      [1999/12/01 v1.4c Il Nuovo Cimento]
\documentclass{cimento}

\usepackage{caption}
\usepackage{amsmath}
\usepackage{graphicx}
\usepackage{epstopdf}
\usepackage{subfigure}
\title{Doubly heavy baryons in a quark model  with AdS/QCD inspired potential}
\author{Floriana~Giannuzzi\from{ins:x}\ETC}
\instlist{\inst{ins:x} University of Bari and INFN - Bari,
       Italy}
\PACSes{\PACSit{12.39.Pn}{Potential models}
\PACSit{14.20.Pn}{Charmed baryons}
\PACSit{14.20.Mr}{Bottom baryons}}

\begin{document}

\maketitle

\begin{abstract}
The spectrum of doubly heavy baryons, hadrons made up of two heavy quarks and one light quark, is computed through a potential model with relativistic kinematics. The expression for the $Q\bar Q$ potential comes from the AdS/QCD correspondence. 
\end{abstract}

In 2002 the Selex Collaboration observed  the baryon $\Xi_{cc}^+ \,(dcc)$ in the decay $\Xi_{cc}^+\rightarrow\Lambda_cK^-\pi^+$ \cite{Mattson:2002vu} and later in $\Xi_{cc}^+\rightarrow pD^+K^- $  \cite{Ocherashvili:2004hi}. The mass of the state was measured to be:
\begin{equation}\label{csiccselex}
M_{\Xi_{cc}}=3518.9 \pm 0.9 \mbox{ MeV .}
\end{equation}
This is the only observed baryon comprising two heavy quarks: no other experiment has observed these states so far. However the quark model predicts their existence and is able to estimate their masses. \\
Baryons can be studied by potential models in two different ways. The first one consists in studying the three body problem, in which the potential term is the sum of the interactions between each pair of quarks. The second one consists in studying the two body problem, in which one quark interacts with a bound state of the other two quarks, in such a way that the problem becomes as simple as studying a meson. 
 In this paper the second approach is followed  \cite{Giannuzzi:2009gh}, supposing that the two heavy quarks are close enough to be seen as one particle (diquark) by the light quark.
The  relativistic kinematics  is introduced by the Salpeter equation for $S$ wave hadrons:
 \begin{equation}\label{salpetereq}
\left(\sqrt{m_1^2-\nabla^2}+\sqrt{m_{2}^2-\nabla^2}+V(r)\right)
\psi({\bf r})\,=\,M\, \psi({\bf r})
\end{equation}
while the potential term results from a recent computation of the quark-antiquark potential within the AdS/QCD correspondence  \cite{Andreev:2006ct}, a recently developed approach to the non perturbative regime of QCD, inspired by the AdS/CFT correspondence. 
%It is obtained computing the expectation value of a rectangular Wilson loop in the AdS space and using the correspondence to relate it to the $q-\bar q$ potential. 
It is  expressed as a parametric equation, in which the energy and the distance between the quark and the antiquark are functions of a parameter $\lambda$ ($\lambda \in [0,2[$):
\begin{equation}  
\label{potadsqcd} \left\{
\hspace{-.4cm}
\begin{array}{cc}
 & V_{AdS/QCD}(\lambda)\,=\,\frac{g}{\pi}
\sqrt{\frac{c}{\lambda}} \left( -1+\int_0^1 dv \, v^{-2} \left[
\mbox{e}^{\lambda v^2/2} \left(1-v^4
\mbox{e}^{\lambda(1-v^2)}\right)^{-1/2}-1\right]\right) \\
& r(\lambda)\,=\,2\, \sqrt{\frac{\lambda}{c}} \int_0^1 dv\, v^{2}
\mbox{e}^{\lambda (1-v^2)/2} \left(1-v^4
\mbox{e}^{\lambda(1-v^2)}\right)^{-1/2}  \hspace{3.3cm} \,.
\end{array}
\right.\end{equation}

%\begin{center}
%\begin{figure}[h]
%\hspace{.8cm}
%\begin{minipage}[c]{.35\textwidth}
%\subfigure{
%\label{}
%\includegraphics[scale=.33]{rlambda}
%}
%%\hspace*{1cm}
%\subfigure{
%\label{}
%\includegraphics[scale=0.4]{vlambda}
%}
%\end{minipage}%
%\begin{minipage}[c]{.5\textwidth}
%\subfigure{
%\label{}
%\includegraphics[scale=0.8]{parpotential}
%}
%\end{minipage}
%\caption{}
%\end{figure}
%\end{center}

The use of this potential term represents the main novelty of this computation. For comparison, one can consider the results found by other potential models  \cite{Valcarce:2008dr,Roberts:2007ni,Ebert:2002ig,Kiselev:2001fw}, QCD Sum Rules \cite{Zhang:2008rt}  or  lattice QCD  \cite{Mathur:2002ce}. 

A spin interaction is also considered, having the form:
\begin{equation}\label{potspin}
V_{spin}(r)\,=\,A \frac{\tilde\delta(r)}{m_1 m_2}{\bf S_1}\cdot{\bf
S_2} \qquad\quad\mbox{with }\qquad \tilde\delta(r)=\left(\frac{\sigma}{\sqrt{\pi}}\right)^3
e^{-\sigma^2 r^2}\,, 
\end{equation}
with $A$ and $\sigma$  parameters. 
%dire perche quark pesanti 
To avoid a divergence in the wavefunction at $r=0$, coming from the divergent potential, a cutoff is added, in order to make the potential constant at small distances, namely for $r<r_M=4\pi\Lambda/(3M)$ ($\Lambda=1$ if $m_1=m_2$). 
The Salpeter equation \eqref{salpetereq} is solved using a numerical method, the Multhopp method \cite{Colangelo:1990rv}.  All  the parameters  are fixed fitting the spectrum of mesons comprising two heavy quarks or one heavy and one light quark. 
The parameters that better reproduce meson spectrum are \cite{Giannuzzi}:  $c=$0.4 GeV$^2$, $g$=2.50, $V_0$=-0.47 GeV (constant term) for the AdS/QCD inspired potential  \eqref{potadsqcd};  $\sigma$=0.47 GeV, $A_c$=14.56, $A_b$=6.49 for the spin term  \eqref{potspin} (two different parameters have been introduced to describe the spin splitting of mesons with charm and bottom quark); $\Lambda=0.5$;  the constituent quark masses $m_q$=0.34 GeV ($q=u,d$), $m_s$=0.48 GeV, $m_c$=1.59 GeV and $m_b$=5.02 GeV.

Since baryons are studied as bound states of a light quark and a heavy diquark, it is necessary to start  computing diquark masses. They are obtained solving the Salpeter equation (\ref{salpetereq}) for the interaction of two heavy quarks: $m_1$ e $m_2$ are the quark masses, $\psi$ and $M$ are the wavefunction and the mass of the diquark, respectively, and $V(r)$ is the potential of interaction of two quarks. In the one-gluon-exchange approximation two quarks can attract each other forming a diquark in the $\bar 3$ representation of $SU(3)_c$ and the energy of the interaction is one half of the quark-antiquark one. So, $V(r)$ in the case of diquarks is one half of the sum of  \eqref{potadsqcd} and \eqref{potspin}. This fixes diquark masses; the computed values are reported in Table~\ref{tabdiquark}. 

\renewcommand{\arraystretch}{.5}
\renewcommand{\tabcolsep}{.5mm}
\begin{table}
\begin{minipage}[c]{.47\textwidth}
 \caption{{\footnotesize Diquark masses in GeV.  Curly ({\it resp.} squared) brackets indicate a spin 1 ({\it resp.} spin 0)
diquark; $n$ is the radial number.} }
\label{tabdiquark}
{\scriptsize
\begin{tabular}{rcl}
\hline
Diquark & State & Mass \\
\hline 
$\{cc\}_{nS}$&1$S$ & 3.238 \\
& 2$S$ & 3.589\\
\hline $[bc]_{nS}$& 1$S$ &6.558\\
& 2$S$ & 6.882 \\
\hline $\{bc\}_{nS}$&1$S$ & 6.562  \\
& 2$S$ & 6.883 \\
\hline $\{bb\}_{nS}$&1$S$ &9.871 \\
& 2$S$ & 10.165\\
\hline
\end{tabular}
}
\end{minipage}%
~\hfill~
\begin{minipage}[c]{.50\textwidth}
\begin{center}
\includegraphics[width=5.5cm]{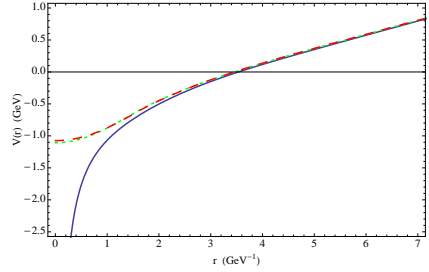}
\captionof {figure}{{\footnotesize Full $Q\bar Q$ potential from Eqs. \eqref{potadsqcd},\eqref{potspin} (solid line); modified potential in Eq. \eqref{barpot} through the wavefunction of $\{cc\}_{1S}$ (dashed line) and $\{cc\}_{2S}$ (dot-dashed line).}}
\label{figmodpotential}
\end{center}
\end{minipage}
\end{table}

Diquark masses are used as input to compute baryon masses. The starting point is again Eq. \eqref{salpetereq}: when applied to baryons, $m_1$ is the mass of the light quark, $m_2$ is the mass of the diquark, $\psi$ and $M$ are the wavefunction and the mass of the baryon, respectively, $V(r)$ is the energy of the interaction between a quark and a diquark. Since the diquark is in the $\bar 3$ representation of the $SU(3)_c$ group, its interaction with a quark is analogous to the one between a quark and an antiquark, the only difference arising from the fact that the diquark is not a pointlike object.  So, the diquark-quark potential is again related to the potential sum of \eqref{potadsqcd} + \eqref{potspin}, but it is modified by an interpolation with the diquark wavefunction to account for its  finite size (Fig. \ref{figmodpotential}):
 \begin{equation}
 \tilde V(R)=\frac1{N}\int d{\bf r} \; |\psi_{d}({\bf r})|^2
V( |{\bf R}+{\bf r}|)
\label{barpot}
\end{equation}
with $N$ a normalization constant.

The spectra of  baryons comprising  two charm, and  two bottom  quarks are reported in Tables \ref{tabbarcc} and \ref{tabbarbb}, respectively.  The mass of $\Xi_{cc}$ predicted in this paper  is in agreement with the experimental value \eqref{csiccselex}, assuming that the uncertainty of the evaluation is of the same order as the difference between the experimental and the theoretical value of meson masses in \cite{Giannuzzi}.

\renewcommand{\arraystretch}{.5}
\renewcommand{\tabcolsep}{.5mm}
\begin{table}[h]
\begin{minipage}[c]{.48\textwidth}
\caption{{\footnotesize Masses (GeV) of baryons comprising a diquark  $\{cc\}_{1S}$ and a light quark ($q$ or $s$).} }
\label{tabbarcc}
{\scriptsize
\begin{tabular}{rcccccl}
\hline
Particle & State & $J^P$ & q-d content & Results \\
\hline
$\Xi_{cc}$ & 1$S$&  $\frac{1}{2}^+$&$q\{cc\}_{1S}$ & 3.547    \\
& 2$S$&&& 4.183  \\
& 3$S$&&& 4.640 \\
\hline
$\Xi_{cc}^*$  & 1$S$& $\frac{3}{2}^+$&  $q\{cc\}_{1S} $ &3.719  \\
& 2$S$&&& 4.282\\
&3$S$&& &4.719 \\
\hline
$\Omega_{cc}$  & 1$S$ & $\frac{1}{2}^+$ &$s \{cc\}_{1S}$  & 3.648   \\
& 2$S$&&& 4.268\\
 &3$S$&&& 4.714 \\
\hline
$\Omega_{cc}^*$  &  1$S$ & $\frac{3}{2}^+$ & $s\{cc\}_{1S}$ & 3.770  \\
& 2$S$&&& 4.334 \\  
& 3$S$&&& 4.766 \\
\hline
\end{tabular}
}
\end{minipage}%
\hspace{.4cm}
\begin{minipage}[c]{.49\textwidth}
\caption{{\footnotesize Masses (GeV) of baryons comprising a diquark $\{bb\}_{1S}$ and a light quark ($q$ or $s$).}}
\label{tabbarbb}
{\scriptsize
\begin{tabular}{rccccc|}
\hline
Particle & State & $J^P$& q-d content & Results\\
\hline
$\Xi_{bb}$ & 1$S$& $\frac{1}{2}^+$&$q \{bb\}_{1S}$  &  10.185   \\
& 2$S$&&& 10.751\\
& 3$S$&&& 11.170 \\
\hline
$\Xi_{bb}^*$ & 1$S$ & $\frac{3}{2}^+$ &$q \{bb\}_{1S}$ & 10.216  \\
& 2$S$&&& 10.770 \\
& 3$S$&&& 11.184 \\
\hline
$\Omega_{bb}$ & 1$S$ & $\frac{1}{2}^+$& $s\{bb\}_{1S}$ &10.271  \\
& 2$S$&&& 10.830\\
& 3$S$&&& 11.240 \\
\hline
$\Omega_{bb}^*$ & 1$S$ & $\frac{3}{2}^+$ & $s\{bb\}_{1S}$ &  10.289  \\
& 2$S$&&& 10.839 \\
& 3$S$&&& 11.247 \\
\hline
\end{tabular}
}
\end{minipage}
\end{table}

%\renewcommand{\arraystretch}{.6}
%\renewcommand{\tabcolsep}{.5mm}
%\begin{table}[h]
%\caption{camb Masses (GeV) of baryons composed by a diquark $bc$ in the lowest mass configuration and a light quark ($q$ or $s$).}
%\label{tabbarbc}
%{\scriptsize
%\begin{tabular}{rccccccccccc|}
%\hline
%Particle & State & $J^P$&Quark-diquark content & This paper   &\cite{Roberts:2007ni} &  \cite{Albertus:2006ya}&\cite{Ebert:2002ig}&\cite{Kiselev:2001fw} & \cite{Zhang:2008rt} \\
%\hline
%$\Xi_{bc}$ & 1$S$ & $\frac{1}{2}^+$ &$q\{bc\}_{1S}$ & 6.904  & 7.011 & 6.919 & 6.933 & 6.82 & 6.75   \\
%& 2$S$&&& 7.478  &&&&&\\
%& 3$S$&&& 7.904 &&&&&\\
%\hline
%$\Xi_{bc}'$ & 1$S$ & $\frac{1}{2}^+$ &$q[bc]_{1S}$& 6.920   & 7.047 & 6.948 & 6.963 & 6.85 & 6.95 \\
%& 2$S$&&& 7.485  &&&&&\\
%& 3$S$&&& 7.908 &&&&&\\
%\hline
%$\Xi_{bc}^*$ & 1$S$& $\frac{3}{2}^+$&$q\{bc\}_{1S}$& 6.936   & 7.074 & 6.986 & 6.980 & 6.90 & 8.00  \\
%& 2$S$&&& 7.495  &&&&&\\
%& 3$S$&&& 7.917 &&&&&\\
%\hline
%$\Omega_{bc}$ & 1$S$& $\frac{1}{2}^+$ & $s\{bc\}_{1S}$ &  6.994   & 7.136 & 6.986 & 7.088 & 6.91& 7.02  \\
%& 2$S$&&& 7.559  &&&&&\\
%& 3$S$&&& 7.976 &&&&&\\
%\hline
%$\Omega_{bc}'$ & 1$S$ & $\frac{1}{2}^+$ & $s[bc]_{1S}$ & 7.005   & 7.165 & 7.009& 7.116 & 6.93 & 7.02  \\
%& 2$S$&&& 7.563  &&&&&\\
%& 3$S$&&& 7.977 &&&&&\\
%\hline
%$\Omega_{bc}^* $& 1$S$ & $\frac{3}{2}^+$ &$s\{bc\}_{1S}$ &  7.017   & 7.187 & 7.046& 7.130 & 6.99 & 7.54  \\
%& 2$S$&&& 7.571  &&&&&\\
%& 3$S$&&& 7.985 &&&&&\\
%\hline
%\end{tabular}
%}
%\end{table}%

The excited states of doubly heavy baryons can also be evaluated  considering the interaction of a $2S$ diquark with a quark. The energy level of this configuration can be compared with the energy of the first radial resonance (2$S$ state) of the baryon comprising the same constituent quarks. The first level is higher than the second for baryons comprising one heavy quark, while it becomes lower for baryons comprising two heavy quarks \cite{Roberts:2007ni}. The masses of the bound states of $2S$ diquark and a quark are reported in Table \ref{bar2sd}: the obtained values are comparable with the ones found with other models. 

\renewcommand{\arraystretch}{.5}
\renewcommand{\tabcolsep}{.5mm}
\begin{table}[h]
\caption{\footnotesize Masses (GeV) of  baryons comprising a diquark in the 2$S$ state and comparisons with the results of other models.}
{\scriptsize
\begin{tabular}{rccccccc|}
\hline
Baryon & $J^P$& q-d content& This paper & \cite{Roberts:2007ni} & \cite{Ebert:2002ig}&  \cite{Kiselev:2001fw} \\
\hline
  $\Xi_{cc}$ & $\frac{1}{2}^+$& $q\{cc\}_{2S}$& 3.893 & 4.029 &  3.910&  3.812\\
 \hline
 $\Xi_{cc}^*$& $\frac{3}{2}^+$&$q\{cc\}_{2S}$&4.021& 4.042 & 4.027& 3.944\\
 \hline
 $\Omega_{cc}$& $\frac{1}{2}^+$&$s\{cc\}_{2S}$& 3.992&  4.180 & 4.075& \\
 \hline
 $\Omega_{cc}^*$& $\frac{3}{2}^+$&s$\{cc\}_{2S}$& 4.105&  4.188  & 4.174& \\
 \hline
 $\Xi_{bb}$& $\frac{1}{2}^+$& $q\{bb\}_{2S}$&10.453&  10.576 &  10.441& 10.373\\
 \hline
 $\Xi_{bb}^*$& $\frac{3}{2}^+$& $q\{bb\}_{2S}$& 10.478& 10.578 &  10.482& 10.413\\
 \hline
 $\Omega_{bb}$& $\frac{1}{2}^+$&$s\{bb\}_{2S}$& 10.538& 10.693 & 10.610& \\
 \hline
 $\Omega_{bb}^*$ & $\frac{3}{2}^+$& $s\{bb\}_{2S}$&  10.556 & 10.721 &  10.645 & \\
  \hline
\end{tabular}
}
\label{bar2sd}
\end{table}%

The results presented here can be analysed through HQET, supposing that a $1/m_{\{QQ\}}$ expansion can be performed for the mass of doubly heavy baryons, in analogy with \cite{Jenkins:1996de}:
\begin{equation}\label{hqetbaryon}
M_{\{QQ\}q}=m_{\{QQ\}}+\bar \Lambda+\frac{\lambda_1}{2m_{\{QQ\}}}+A_Q d_H \frac{\lambda_2}{2 m_{\{QQ\}}}
\end{equation}
$m_{\{QQ\}}$ being the mass of the diquark and  $d_H={\bf S}_{\{QQ\}}\cdot {\bf S}_q$. From \eqref{hqetbaryon}  the mass splitting between baryons with $J^P=3/2^+$ and  $J^P=1/2^+$ can be predicted. These predictions are verified by the results found in this paper. 

The comparison with the predictions of HQET could be important for testing   the validity of the model. However the proof of the existence of doubly heavy baryons  can only come from the experimental observations, for which we have to wait for forthcoming experiments.

\acknowledgments
I thank P.~Colangelo, F.~De Fazio and S.~Nicotri  for precious suggestions. I also thank M.V. Carlucci, M. Pellicoro and S. Stramaglia  for collaboration on developing the numerical method used here. This work was supported in part by the EU Contract No. MRTN-CT-2006-035482, "FLAVIAnet".

\end{document}